\documentclass[twocolumn]{aastex702}
\usepackage{newtxtext}
\usepackage[slantedGreek]{newtxmath}
\usepackage{rnaas-one}
\usepackage{attachfile}
\journalinfo{RNAAS}
\begin{document}

\title{Arcminute Microkelvin Imager observations at 15.5~GHz of the
tidal disruption event Swift J164449.3$+$573451 from 2011 Mar to 2014
Mar}

\author[0000-0003-3189-9998]{David A.\ Green}
\affiliation{Cavendish Laboratory, University of Cambridge,\\
   J.~J.\ Thomson Ave., Cambridge, CB3 0US, UK}
\email{dag9@cam.ac.uk}

\begin{abstract}
Radio monitoring of the tidal disruption event Swift J164449.3$+$573451
at 15.5~GHz with the Arcminute Microkelvin Imager is reported. The
results from 308 observations made from 2011 Mar to 2014 Mar are
presented. These show a steady rise in flux density from $\approx 3$~mJy
(three days after the initial $\gamma$-ray detection of Swift
J164449.3$+$573451), to a peak of nearly $30$~mJy $\approx 140$~days
later, followed by a steady decline.
\end{abstract}


\section{Introduction}

Swift J164449.3$+$573451 is a tidal disruption event (TDE), first
detected in $\gamma$-rays on 2011 Mar 28 by the Neil Gehrels Swift
Observatory \citep{2011GCN.11823....1C}. It was originally designated as
a $\gamma$-ray burst (GRB 110328A), but as it re-triggered Swift within
an hour it was noted that it was either an usually long GRB, or some
other sort of transient \citep{2011GCN.11824....1B}. Subsequently it was
identified as a TDE \citep[{e.g.}][]{2011Sci...333..203B,
2011Natur.476..421B, 2011Natur.476..425Z}, of a star onto the nuclear
black hole of a galaxy, in this case at a redshift $z \approx 0.35$
\citep{2011GCN.11833....1L}. In X-rays it showed short time scale dips
in intensity, with an overall steady decline, to $\approx 500$ days
after its initial detection, when it showed a sharp drop in intensity
\citep[{e.g.}][]{2012MNRAS.422.1625S, 2013ApJ...767..152Z}. At radio
wavelengths Swift J164449.3$+$573451 brightened for the first hundred
days or so, depending on frequency, after which it showed a steady
decline \citep{2012ApJ...748...36B, 2013ApJ...767..152Z}.

Here radio monitoring of Swift J164449.3$+$573451 from 2011 Mar to 2014
Mar with the Arcminute Microkelvin Imager
\citep[AMI,][]{2008MNRAS.391.1545Z} are reported. These include
observations from 2011 Mar to 2012 Nov, the results from which have
already been already presented in \citet{2012ApJ...748...36B} and
\citet{2013ApJ...767..152Z}. Here these observations have been
reprocessed, along additional observations from 2012 Nov to 2014 Mar,
giving coverage the first three years of the Swift J164449.3$+$573451
outburst.

\section{Observations}

The observations were made with the AMI `Large Array' which is a radio
interferometer consisting of eight 13.5-m diameter antennas. A single
linear polarisation, Stokes parameter $I+Q$, was observed over a
frequency range of 13 to 18~GHz. The observations were made using an
analog correlator, which split the observed band into 8 broad frequency
channels. The end channels have lower sensitivity and are also prone to
interference, so the observations presented here exclude the end
channels.

\begin{figure*}[ht!]
\centerline{\includegraphics[clip=,width=16.5cm]{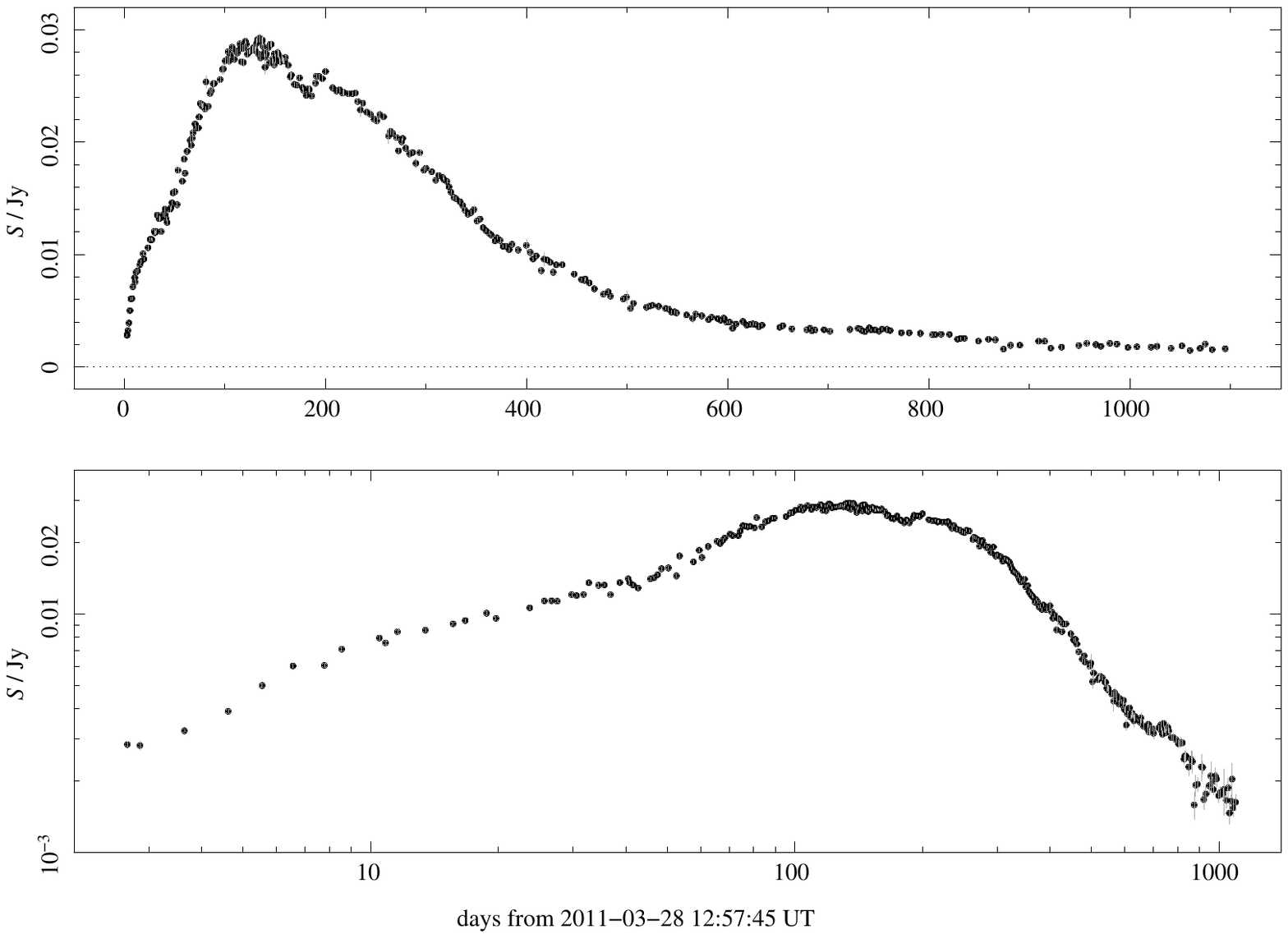}}
\caption{Radio light curve of Swift J164449.3$+$573451 at 15.5~GHz from
2011 Mar to 2014 Mar from AMI observations, plotted linearly (top panel)
and logarithmically (bottom panel). Each data point is a the average
from an observation. Statistical error bars are plotted, although these
are usually smaller than the size of symbols.\label{fig:ami-la}}
\end{figure*}

The observations usually consisted of multiple 10-min scans of Swift
J164449.3$+$573451 interleaved with 100-s observations of a nearby,
compact calibrator source J1638$+$5720. Observations, varied in length,
typically with at least two 10 min scans on source, but sometimes longer
observations, up to several hours, were made. Swift J164449.3$+$573451
was first observed on 2011 Mar 31, 02:07 to 08:06 UT
\citep{2011GCN.11849....1P}. Initially observations were made daily,
with later observations spaced two or more days apart. Some observations
are excluded from the subsequent analysis, due to bad weather (either
high winds requiring the antennas to be stowed, or heavy rain, which
makes accurate calibration not possible), or technical issues. The
results from \textattachfile[color=0.2 0.6 0.2]{AMI-swift-tde.txt}{308 observations}, from
2011 Mar 31 to 2014 Mar 27, are presented here.

\section{Analysis and Results}

The data were processed using standard procedures, using the
\texttt{reduce} package. The flux density scale established from short
observations of the standard calibrator sources 3C286 or 3C48 which were
made on most days, including the `rain gauge' measurements made during
the observations reflecting different atmospheric conditions (see
\citeauthor{2008MNRAS.391.1545Z}). The data were flagged: (i)
automatically to eliminate bad data due to various technical problems
and interference; (ii) manually, to eliminate remaining interference and
some periods with heavy rain during longer observations. The interleaved
observations of J1638$+$5720 provided: (i) the phase calibration of each
antenna in the array throughout each observation, and (ii) adjusted the
observation-to-observation flux density scale, based on a smooth
(spline) fit to the apparent flux density of J1638$+$5720 (which varied
slowly, from $~0.8$ to 2.0~Jy, with an {r.m.s.} scatter of $\approx 5$\%
of the apparent flux densities from the spline fit).

Figure~\ref{fig:ami-la} shows the 15.5-GHz light curve for Swift
J164449.3$+$573451 from these observations. Flux densities were derived
for each observation, from 6 broad frequency channels covering 13.6 to
17.4~GHz, and then a power law fit was made to obtain a flux density at
15.5~GHz. The results presented here have smaller statistical errors
than the earlier processing of these observations presented in
\citet{2012ApJ...748...36B} and \citet{ 2013ApJ...767..152Z}.

\begin{acknowledgments}
All the AMI observations of Swift J164449.3$+$573451 -- which are
reanalysed here -- were made by my former colleague Guy Pooley, who died
in 2020. I acknowledge his contribution with gratitude. I also thank the
staff of the Mullard Radio Astronomy Observatory, University of
Cambridge, for their support in the maintenance, and operation of AMI.
\end{acknowledgments}

\end{document}